\documentclass[nofootinbib,aps,pra,twocolumn,superscriptaddress,floatfix]{revtex4-2}
\usepackage[english]{babel}
\usepackage[utf8]{inputenc}
\usepackage[T2A]{fontenc}
\usepackage{hyperref, xcolor, graphicx}
\usepackage{amsfonts, amssymb, amsmath, mathtools}
\usepackage{subcaption}
\usepackage{siunitx}
\usepackage{tikz}
\usepackage{calc}
\usepackage{ragged2e}

\usepackage{float}

\hypersetup{
	colorlinks=true,
	linkcolor=black,
	citecolor=blue,   
	urlcolor=blue,
}

\usepackage{bibunits}

\defaultbibliographystyle{unsrt}

\begin{document}

	\title{Optomagnetic non-thermal modification of the ferromagnetic resonance}
	
	\author{Nika Gribova}
	\email{gribova.ni@phystech.edu}
	\affiliation{Russian Quantum Center, Moscow 121205, Russia}
	\affiliation{Moscow Institute of Physics and Technology, Dolgoprudny 141701, Russia}
    \affiliation{Lomonosov Moscow State University, Moscow 119991, Russia}
	\author{Anatoly Zvezdin}
	 \affiliation{Prokhorov General Physics Institute of the Russian Academy of Sciences, Moscow 119991, Russia}
    \affiliation{Russian Quantum Center, Moscow 121205, Russia}
    \affiliation{Lomonosov Moscow State University, Moscow 119991, Russia}
	 \author{Shixun Cao}
	 \affiliation{Department of Physics, Materials Genome Institute, Institute for Quantum Science and Technology, Shanghai University, Shanghai 200444, China}
	\author{Vladimir Belotelov}
	\affiliation{Russian Quantum Center, Moscow 121205, Russia}
	\affiliation{Lomonosov Moscow State University, Moscow 119991, Russia}
	
	\date{\today}
	
	\begin{abstract}
		We investigate the optomagnetic shift of the ferromagnetic resonance (FMR) frequency in magnets caused by the inverse Cotton-Mouton effect (ICME) under linearly polarized light. Using a Lagrangian description of magnetization dynamics, we derive the equations of motion, and obtain analytical expressions for the resonance frequency in both in-plane and out-of-plane equilibrium configurations. The theory shows that the FMR frequency depends on the polarization angle and propagation direction of light, with ICME producing a frequency shift that can dominate over thermal effects. The analytical results agree well with numerical simulations and with available experimental data for bismuth-substituted yttrium iron garnet, enabling estimation of the ICME contribution. These findings demonstrate that linearly polarized light can be used to control ferromagnetic resonance through magneto-optical effects.
	\end{abstract}
	
	\maketitle
	
	\begin{bibunit}

	\section*{Introduction}

Over the past several decades, the interaction between light and spins in magnetically ordered materials has emerged as a focal point of condensed matter research \cite{kimel2005ultrafast, kalashnikova2007impulsive, yoshimine2014phase, stupakiewicz2019selection, krichevsky2024spatially, gribova2026unified}. Initial investigations primarily utilized femtosecond laser pulses to induce both thermal \cite{ostler2012ultrafast, soumah2021optical, frej2023nonlinear} and non-thermal \cite{lutsenko2024magnetophotonic} modifications of spin states. Non-thermal mechanisms are driven by photoinduced magnetic anisotropy \cite{ostler2012ultrafast, stanciu2007all} or optomagnetic phenomena, including the inverse Faraday and Cotton-Mouton effects (IFE and ICME, respectively). These interactions facilitate advanced functionalities such as ultrafast all-optical switching \cite{ostler2012ultrafast, stanciu2007all, vahaplar2009ultrafast, gribova2026optomagnonic, frej2023nonlinear} and the coherent excitation of spin waves \cite{satoh2012directional, savochkin2017generation}.

While transient pulse excitation is prominent, continuous-wavelaser irradiation also offers a viable pathway for spin manipulation. Early research focused on photomagnetic data storage \cite{jiao2017dependence, richter2014heat}, largely relying on localized laser heating to modulate magnetization and anisotropy—a principle now industrialized via Heat-Assisted Magnetic Recording \cite{vogler2016heat, weller2014hamr}. However, such thermal effects necessitate significant optical absorption to drive electronic transitions. In transparent dielectrics, specifically rare-earth iron garnets, absorption-mediated heating is minimized, allowing non-thermal, inverse magneto-optical effects to dominate. These include the Inverse Faraday Effect, driven by circularly polarized light \cite{kimel2005ultrafast, pershan1963nonlinear}, and the Inverse Cotton–Mouton Effect, triggered by linearly polarized light \cite{zon1987observation, popova2011theory, zvezdin2024giant}.

Ferromagnetic resonance lies in roots of many magnetic experiments and applications and is governed by the effective field entering the Kittel \cite{kittel1948theory, kittel1951ferromagnetic} relation and therefore depends on external bias. The FMR in garnet films is modulated by external field magnitude and orientation \cite{lee2016ferromagnetic}, film thickness \cite{liu2022strain, rao2018thickness, ding2020nanometer, krysztofik2021effect}, chemical composition \cite{randoshkin1999characteristic, rosenberg2021magnetic, das2023perpendicular}, and temperature-dependent magnetic properties \cite{jermain2017increased, haidar2015thickness, panin2025exploring, laulicht1991temperature}. Furthermore, tuning is achieved through elastic strain \cite{krysztofik2021effect, ding2020nanometer, deb2018picosecond}, electric-field coupling \cite{yu2019nonvolatile, zavislyak2013electric, zhang2014electric}, and light-induced anisotropy changes ranging from steady-state shifts to ultrafast precessional triggering \cite{ stupakiewicz2001light, soumah2021optical, atoneche2010large}.

Dynamic control of ferromagnetic resonance frequency via external stimuli is a critical objective, with optical modulation offering high precision. The Inverse Cotton-Mouton Effect facilitates non-thermal magnetic anisotropy modification within a material's transparency range. In ferrimagnetic garnets like Y$_3$Fe$_5$O$_{12}$ (YIG) and bismuth-substituted variants (BiYIG), ICME enables efficient magnetization precession triggering via linearly polarized light, often surpassing the efficiency of the inverse Faraday effect or photo-induced anisotropy \cite{shen2018dominant, yoshimine2014phase}. Furthermore, in antiferromagnets such as orthoferrites and NiO, ICME-driven coherent magnon excitation by linearly polarized pulses is orders of magnitude stronger than circular polarization-driven effects \cite{iida2011spectral, kalashnikova2007impulsive}. These findings indicate that ICME can be used for fundamentally shifting FMR frequency through optical modulation of the effective magnetic field.

As the experimental foundation for our research, we employ the results obtained by \cite{polulyakh2022light}, where a key experiment relevant to the present problem was carried out. While the experimental data are of significant value, the theoretical description is not complete. In this work, a consistent and rigorous theory is formulated, providing full agreement with the experimental observations from \cite{polulyakh2022light} and given additional insights into light mediated non-thermal control of FMR.

In this work, we investigate the dependence of the FMR frequency on the in-plane orientation of the polarization of light at room temperature. Experimental data indicate that ICME provides the main contribution to the polarization-dependent frequency shift, that dominates photothermal contribution to shift of FMR frequency. We analyze the observed frequency shifts in terms of anisotropy fields generated by ICME and their relation to static anisotropies.

\section*{Theory}

Let's consider an iron garnet thin film in an external in-plane magnetic field \(\mathbf{H} = (H, 0, 0)\). The Cartesian coordinate system is chosen to get z-axis out-of-plane and x-axis along the external magnetic field (Fig. 1).
The film exhibits uniaxial magnetic anisotropy in the cases of easy-axis, while the effects associated with cubic crystallographic anisotropy are not taken into account. The sample is illuminated by linearly polarized light characterized by the electric field \(\mathbf{E} = E_0 (\sin\beta \cos\alpha, \sin\beta \sin\alpha, \cos\beta)\), which is parameterized by two angles \(\alpha\) and \(\beta\). Here, \(\beta\) denotes the angle of deviation of the polarization of incident light from the normal of the film, while \(\alpha\) represents the rotation of the in-plane polarization projection relative to the \(x\)-axis.

\begin{figure}
\centering
\includegraphics[width=0.8\linewidth]{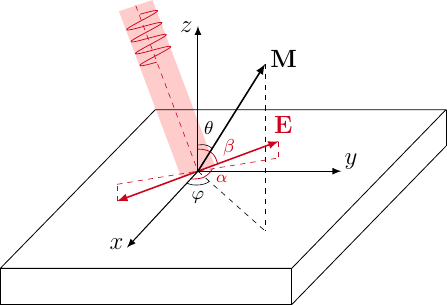}
\caption{\justifying{ 
		Ferromagnetic film with uniaxial magnetic
		anisotropy in the external magnetic field $H_{\mathrm{ext}}$ along $x$
		axis.
	}
}
\label{fig:in_plane}
\end{figure}

To describe the dynamics of magnetization in this system we employ the Lagrangian formalism. 
The spherical coordinates are chosen as the polar and azimuthal angles $(\theta,\varphi)$ of the magnetization $\mathbf{M}=M(\sin\theta\cos\varphi, \sin\theta\sin\varphi, \cos\theta)$ in Cartesian system. 
The Lagrangian of the system is constructed from the kinetic term and the potential energy. 
\begin{equation}
\mathcal{L} = -\frac{M}{\gamma}\dot{\phi}\cos{\theta} - U_a - U_Z - U_d - U_{cm}
\label{}
\end{equation}
The total potential energy $U$ includes several contributions: Zeeman energy $U_z$ due to the static magnetic field, uniaxial magnetocrystalline anisotropy $U_a$, demagnetization energy $U_d$, spin-photon interaction energy described by the inverse Cotton–Mouton effect $U_{cm}$. The exchange energy was omitted, since we consider the film to be in a monodomain state. The above energy contributions can be written in spherical coordinates as
\begin{align}
&U_a = -K_u \cos^2\! \theta,\label{ua}\\
&U_z = 
- M H\sin\theta \cos\varphi,\label{uz}\\
&U_d = 2\pi M^2 \cos^2\theta,\label{ud}\\
&U_{cm} = K_{cm} \Bigl[ \cos2\beta +\cos2\theta + 3\cos2\beta\cos2\theta  \label{ucm}\\
&\qquad +\sin^2\!\beta \sin^2\!\theta\bigr(4\cos2\alpha\cos2\varphi + \sin2\alpha\sin2\varphi\bigr)\nonumber \\
&\qquad+\cos(\alpha-\varphi)\sin2\beta\sin2\theta\Bigr]\nonumber,
\end{align}
where derivation of Cotton-Mouton energy is given in detail in Appendix~\ref{appendix_a}.

Constructing the Euler–Lagrange equations based on the Lagrangian with the energy terms presented in Eqs.~(\ref{ua}-\ref{ucm}), one obtains the equations of magnetization motion for $\theta(t)$ and $\varphi(t)$. 
\begin{align}
	&\dot{\theta} =   -\gamma H \sin\varphi\label{theta}\\
	&\quad  
	-\omega_{cm}\Bigl[\sin\theta\sin^2\!\beta(\sin2\alpha\cos2\varphi-4\cos2\alpha\sin2\varphi)\nonumber\\
    &\quad +\cos\theta\sin2\beta\sin(\alpha-\varphi)\Bigr]\nonumber\\
	&\dot{\varphi} = \cos\theta \Bigr(\omega_u-2\omega_{cm}(1+3\cos2\beta)-\frac{\gamma}{\sin\theta} H \cos\varphi\Bigl)\label{phi}\\
	&\quad  +\omega_{cm}\Bigl[\cos\theta\sin^2\!\beta(4\cos2\alpha\cos2\varphi+\sin2\alpha\sin2\varphi)\nonumber\\
    &\quad +\frac{\cos2\theta}{\sin\theta}\sin2\beta\cos(\alpha-\varphi)
    \Bigl]\nonumber
\end{align}
where $\gamma$ is a gyromagnetic ratio, $\omega_{cm}=\frac{2\gamma K_{cm}}{M}$, $\omega_u = \frac{2\gamma (K_u-2\pi M^2)}{M}$ and $K_{cm} = (a_1-a_2)M^2 E_0^2/8$ with constants $a_1, a_2$, representing two possible ICME energy contributions admitted by the symmetry group of the system (Appendix~\ref{appendix_a}).
These equations describe the precessional dynamics of magnetization determined by uniaxial anisotropy and external fields of a magnetic field and linearly polarized light, when the system is perturbed from an equilibrium position.

The equilibrium state $(\tfrac{\partial U}{\partial \theta} = 0, \tfrac{\partial U}{\partial \varphi} = 0$ with $\omega_{cm}=0)$ is determined by the value of the uniaxial anisotropy and the external magnetic field. Here we will consider in detail the case when initially the magnetization is alongside the external magnetic field: $\theta_0 = \pi/2$, $\varphi_0 = 0$. For the magnetic film with the "easy-axis" anisotropy ($K_u>2\pi M^2$) it is realized if $H > \tfrac{2K_u}{M} - 4\pi M$. If the uniaxial anisotropy of the magnetic film is "easy-plane" type ($K_u<2\pi M^2$) then any external magnetic field along x-axis provides this equilibrium state. The case of the out-of-plane magnetization equilibrium position is described in the Apendix~\ref{appendix_b}.

\begin{figure*}[htb!] 
	\centering
	\includegraphics[width=1\linewidth]{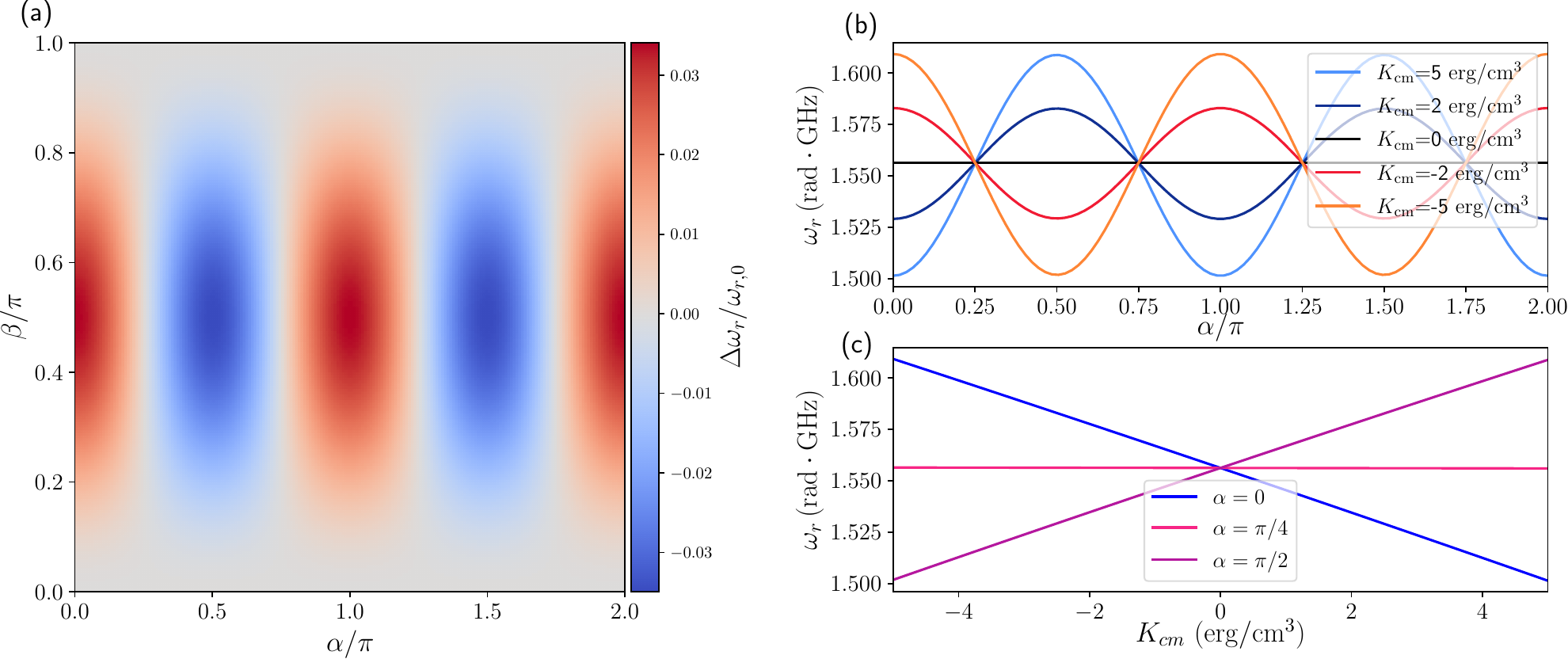}
	\caption{\justifying{ 
			The resonant frequencies dependence on the parameters $\alpha$, $\beta$ and $K_{cm}$ with equilibrium position in-plane with $K_u-2\pi M^2<0$.
			(a) The colormap with fixed parameter $K_{cm}=-5~\mathrm{erg/cm^3}$ illustrates analytical expression for $\Delta\omega_r/\omega_{r, 0}$ derived from the Eq.~\ref{wr}, where $\omega_{r,0}=\omega_r(K_{cm}=0)$ and $\Delta\omega_r=\omega_r-\omega_{r, 0}$. 
			(b) Dependence of $\omega_r$ at normal light incidence ($\beta=\pi/2$) on the polarization angle $\alpha$ for several values of ICME parameter $K_{\mathrm{cm}}$. 
			(c) Dependence of $\omega_r$ at normal light incidence on $K_{cm}$ for $\alpha=0, \pi/4, \pi/2$.
		}
	}
	\label{fig:in_plane}
\end{figure*}

Therefore, to linearize Eqs.~(\ref{theta}-\ref{phi}) it is convenient to pass small angles $\theta_1,\varphi_1 \ll 1$ as $\varphi = \varphi_0+\varphi_1$ and $\theta = \theta_0+\theta_1$. Taking into account terms up to the first order in $\theta_1, \phi_1$, as numerical modeling indicates that this is sufficient to accurately describe the system, we get the following linearized equation for the free magnetization precession
\begin{align}
	\dot{\theta}_1 = -\omega_1\varphi_1+\omega_3\theta_1+\omega_5,\label{linear_theta1}\\
	\dot{\varphi}_1 = \omega_2\theta_1+\omega_4\varphi_1 +\omega_6,\label{linear_varphi1}
\end{align}
where $\omega_1 = \gamma H -8\omega_{cm}\cos2\alpha\sin^2\!\beta$, $\omega_2 = \gamma H - \omega_{u}+2\omega_{cm}(1+3\cos2\beta-2\cos2\alpha\sin^2\!\beta)$, $\omega_3 =-\omega_4= \omega_{cm}\sin\alpha\sin2\beta$, $\omega_5 = -\omega_{cm}\sin2\alpha\sin^2\beta$ and $\omega_6 = -\omega_{cm}\sin2\beta\cos\alpha$. These linearized equations \ref{linear_theta1}, \ref{linear_varphi1} can be solved exactly under the initial conditions $\theta_1\!(0)$ and $\varphi_1\!(0)$, assuming $\frac{\omega_{cm}}{\gamma H_x}\ll1$ we have
\begin{align}
	&\varphi_1(t) = \frac{1}{\omega_r^2}\Bigl[\Omega_1^2 + \Omega_2^2\cos\omega_rt+\omega_r\Omega_3\sin\omega_rt\Bigr],\label{in_exact_phi}\\
	&\theta_1(t) = \frac{1}{\omega_r^2}\Bigl[\Omega_4^2 + \Omega_5^2\cos\omega_rt+\omega_r\Omega_6\sin\omega_rt\Bigr],\label{in_exact_theta}\\
	&\omega_r=\sqrt{\omega_1\omega_2-\omega_3^2}\label{wr},
\end{align}
where $\omega_r$ is the resonant frequency, $\Omega_1^2 = \omega_2\omega_5-\omega_3\omega_6$ and  $\Omega_4^2 = \omega_3\omega_5-\omega_1\omega_6$ are determined only by the parameters of the system, $\Omega_2^2=\varphi_1\!(0)\omega_r^2-\Omega_1^2$, $\Omega_3 = \theta_1\!(0)\omega_2-\varphi_1\!(0)\omega_3+\omega_6$, $\Omega_5^2 = \theta_1\!(0)\omega_r^2-\Omega_4^2$ and $\theta_1\!(0)\omega_3-\varphi_1\!(0)\omega_1+\omega_5$ are determined by the system and the initial conditions of magnetization precession $\theta_1\!(0)$ and $\varphi_1\!(0)$. The magnitude and polarization of a linearly polarized electric field affects not only the magnitude of the Cotton–Mouton effect but also induces a frequency shift. The Eq.~\ref{wr} is in good agreement with Kittel formula (Appendix~\ref{appendix_fmr})

In the case of normal incidence of light ($\beta=\pi/2$) the Eq.~\ref{wr} is simplified ($\omega_3=\omega_4=\omega_6=0$) and resonant frequency becomes $\omega_r=\sqrt{\omega_1\omega_2}$.

\section*{Properties of the optomagnetic shift of FMR}

Fig.~\ref{fig:in_plane} demonstrates dependence on $\alpha$, $\beta$ and $K_{cm}$.
The parameters of system are selected for bismuth substituted iron garnet film $\mathrm{Bi Y_2 Fe_{4.4}Sc_{0.6} O_{12}}$ \cite{polulyakh2022light}: $H=8$~Oe, $4\pi M = 1830~$Oe, $\gamma =1.76 \times 10^{-5}~\text{ps}^{-1}\text{Oe}^{-1}$ and $K_u=61.6\times 10^{3}~$erg cm$^{-3}$.
For this set of parameters $K_u-2\pi M^2 = -70.6\times 10^{3}~$erg cm$^{-3}$ the in plane equilibrium condition is satisfied.

Fig.~\ref{fig:in_plane}(a) presents the relative resonant frequency change $\Delta\omega_r/\omega_{r, 0}$ calculated by Eq.~\ref{wr}, where $\omega_{r,0}=\omega_r(K_{cm}=0)$ and $\Delta\omega_r=\omega_r-\omega_{r, 0}$.
The largest influence of light on FMR appears for normal incidence ($\beta=\pi/2$).

The dependence of $\omega_r$ on the polarization angle $\alpha$ at normal light incidence is plotted in Fig.~\ref{fig:in_plane}(b) for several values of ICME parameter $K_{\mathrm{cm}}$. 
For selected parameters, the peak values of $\omega_r$ correspond to the angles around $\alpha=0, \pi/2, \pi, 3\pi/2$, where polarization of light is either parallel or perpendicular to the equilibrium position of magnetization. On the contrary, if $\alpha$ is nearly $\pi/4, 3\pi/4, 5\pi/4, 7\pi/4$, then there is no change in FMR frequency and $\omega_r=\omega_{r, 0}$.

\begin{figure}[h!]
	\centering
	\includegraphics[width=1\linewidth]{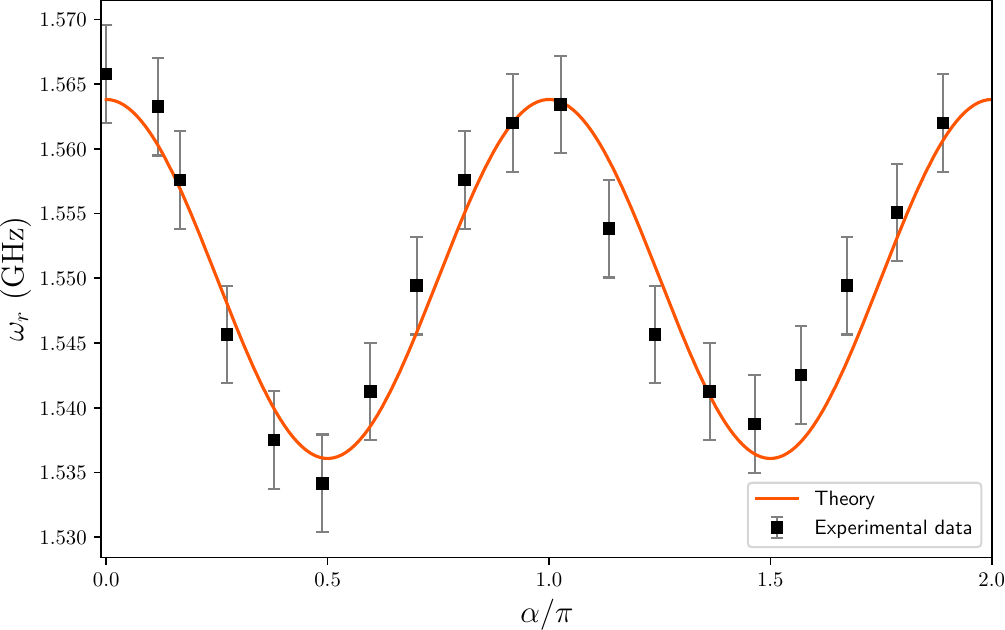}
	\caption{\justifying{Dependence of the FMR frequency on the polarization direction $\alpha$ of the light wave at $T=300\,\mathrm{K}$. A constant magnetic field $H_x=8\,\mathrm{Oe}$ is applied. The light beam with power $P=25\,\mathrm{mW}$ makes an angle of $\approx 5^{\circ}$ with the normal to the sample plane. Experimental data points (black dots) {and error bars} are taken from article \cite{polulyakh2022light}. The solid theoretical curve is calculated according to equation \ref{wr} using the experimental parameters. 
		}
	}
	\label{fig:exp1}
\end{figure}

Larger $K_{cm}$ produces a stronger shift of FMR.
Fig.~\ref{fig:in_plane}(c) demonstrates that the effect gets stronger for increasing $K_{cm}$ and the frequency shift is almost linear in $K_{cm}$, i.e. in the incident light intensity.

In this section we apply the developed theory to describe the experimental results obtained in \cite{polulyakh2022light} where dependence of $\omega_r$ on light polarization was measured.	{
Optical pumping was performed using a continuous-wave laser at a wavelength of $\lambda = 680$ nm with a beam spot size of approximately 6~mm$^2$, which illuminated roughly 24\% of the total surface of the $5 \times 5$ mm$^2$ sample. The static and alternating microwave magnetic fields were aligned orthogonal to each other within the sample plane. To optimize the signal-to-noise ratio, each spectral trace was obtained by averaging 50 consecutive scans. The experimental investigations were conducted on a bismuth-substituted yttrium iron garnet ($\text{BiY}_2\text{Fe}_{4.4}\text{Sc}_{0.6}\text{O}_{12}$) film with a thickness of $d = 12$~$\mu$m, synthesized via liquid-phase epitaxy on a (111)-oriented gadolinium gallium garnet (GGG) substrate \cite{polulyakh2022light}.}

Fig.~\ref{fig:exp1} presents a
theoretical description of the experimental data (black dots) by Eq.~(\ref{wr}) (solid curve). The magnetic parameters are taken the same as in the beginning of section. The best correspondence is achieved for $K_{cm} = -1.25~$erg cm$^{-3}$. Since in the experiment light beam power was 25~mW one can find $a_1-a_2 = -3.1\times10^{-7}/\mathrm{Oe^2}$, which value is the same order as presented in \cite{pisarev1971magnetic}.

\section*{Field-Swept Resonance Field Shift Induced by ICME}
{
To analyze the impact of the inverse Cotton-Mouton effect under standard field-swept ferromagnetic resonance conditions, the microwave driving frequency $\omega$ is treated as a strict constant. In the unperturbed regime, defined by the absence of optical pumping ($\omega_{cm} = 0$), the resonance condition is satisfied precisely at the baseline magnetic field $H = H_r$. Linearizing the unpumped equations of motion around the saturated in-plane equilibrium configuration $(\theta_0, \varphi_0) = (\pi/2, 0)$ yields the unperturbed dispersion relation:
\begin{equation}
    \omega^2 = \gamma H_r(\gamma H_r - \omega_u).
\end{equation}

When the system is optically pumped, the spin-photon coupling ($\omega_{cm} \neq 0$) acts as an effective opto-magnetic anisotropy that modifies the free energy curvature. Introducing small-signal harmonic perturbations $\delta\theta, \delta\varphi \propto e^{i\omega t}$ for the magnetization variations, the dynamics are governed by the coupled linear system: $i\omega\delta\theta = A\delta\theta + B\delta\varphi$ and $i\omega\delta\varphi = C\delta\theta + D\delta\varphi$, where $A = \partial\dot{\theta}/\partial\theta$, $B = \partial\dot{\theta}/\partial\varphi$, $C = \partial\dot{\varphi}/\partial\theta$ and $D = \partial\dot{\varphi}/\partial\varphi$ are as follows  
\begin{align}
    A &= \omega_{cm}\sin2\beta\sin\alpha=-D,\\
    B &= -\gamma (H_r+\Delta H_r) + 8\omega_{cm}\sin^2\beta\cos2\alpha,\\
    C &= \gamma (H_r+\Delta H_r) - \omega_u \\
    &\quad + 2\omega_{cm}(1+3\cos2\beta - 2\sin^2\beta\cos2\alpha).\nonumber
\end{align}
To maintain the fixed-frequency resonance condition ($\delta\omega^2 = 0$) the resonant external magnetic field is taking the form $H=H_r+\Delta H_r$, where $\Delta H_r$ represents the exact field displacement required to compensate for the optomagnetic ICME $\omega_{cm} \neq 0$.

Non-trivial solutions for the coefficients exist only if the determinant of the coefficients vanishes, generating the secular equation
$\omega^2 = -A^2 - BC$. The parameter $A$ is proportional to the spin-photon coupling $\omega_{cm}$. Consequently, the term $A^2 \propto \omega_{cm}^2$ constitutes a second-order perturbation and is neglected in the weak optomagnetic coupling regime ($\omega_{cm} \ll \gamma H, \omega_u$). The resonance condition simplifies to
\begin{align}
    \omega^2 \approx \gamma H_r & (\gamma H_r - \omega_u) + \gamma \Delta H_r(2\gamma H_r - \omega_u) \\
    &+ 2\omega_{cm}\gamma H_r(1+3\cos2\beta - 2\sin^2\beta\cos2\alpha)\nonumber
     \\
     &- 8\omega_{cm}\sin^2\beta\cos2\alpha(\gamma H_r - \omega_u). \nonumber
\end{align}

Subtracting the unperturbed system $\omega^2 = \gamma H_r(\gamma H_r - \omega_u)$ requires the sum of all first-order perturbation terms to strictly vanish. Isolating $\Delta H_r$ provides the explicit analytical expression for the optomagnetic resonance field shift:
\begin{align}
    \Delta H_r(\alpha, \beta) = \frac{2\omega_{cm}}{\gamma (2\gamma H_r - \omega_u)} &\big( - \gamma H_r(1+3\cos2\beta) \\
    &
    +2\sin^2\beta\cos2\alpha(3\gamma H_r - 2\omega_u)
     \big).\nonumber
\end{align}

In the configuration corresponding to the normal incidence of light relative to the sample plane, the polar polarization angle simplifies to $\beta = \pi/2$. Under this geometric condition, one can reduce the general expression for the optomagnetic resonance field shift to the following relation:
\begin{equation}
    \Delta H_r(\alpha) = \frac{4\omega_{cm}}{\gamma} \left[ \frac{\gamma H_r + \cos2\alpha(3\gamma H_r - 2\omega_u)}{2\gamma H_r - \omega_u} \right]
\end{equation}

To analyze the evolution of this opto-magnetic response across distinct magnetic operating regimes, the dependence of $\Delta H_r$ on the azimuthal polarization angle $\alpha$ is evaluated for a wide range of unperturbed baseline fields: $H_r = 8, 100, 500,$ and $2000$ Oe. The simplified expression demonstrates that the field shift retains a strict $\pi$-periodicity dictated by the $\cos2\alpha$ symmetry of ICME (Figure~\ref{fig:Delta_Hr}). Furthermore, the expression reveals that both the vertical offset and the harmonic amplitude of the modulation are fundamentally governed by the ratio of the external Zeeman energy $\gamma H_r$ to the effective uniaxial anisotropy frequency $\omega_u$. As $H_r$ increases far beyond the anisotropy field limit ($H_r \gg \omega_u/\gamma$), the fraction within the brackets asymptotically approaches $(1 + 3\cos2\alpha)/2$, causing the shape of the polarization dependence to transition towards a high-field invariant profile.

\begin{figure}[h!]
	\centering
	\includegraphics[width=1\linewidth]{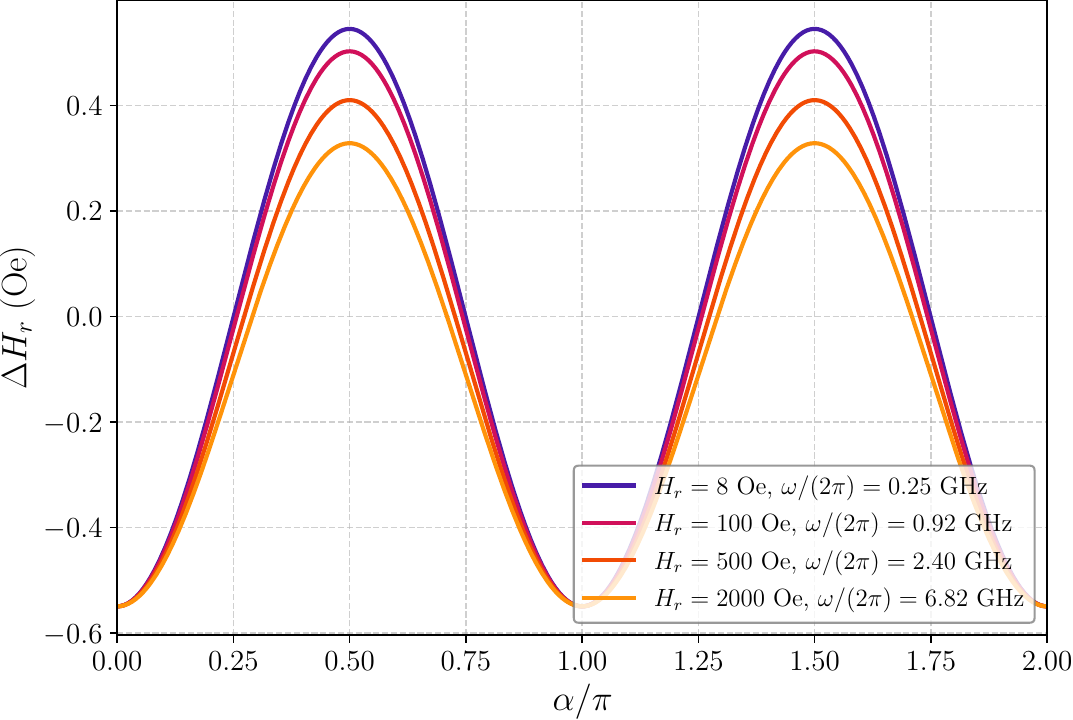}
	\caption{\justifying{
    Azimuthal polarization dependence of the photoinduced resonance field shift $\Delta H_r$ evaluated at a fixed polar angle $\beta = \pi/2$. The continuous curves track the field shift profile across $\alpha/\pi \in [0, 2]$ for three discrete unperturbed resonance fields: $H_r =8, 100,500,2000$~Oe. Fixed material constants utilized in the calculation are $M = 1830/(4\pi)$ G, $K_u = (62.619 \times 10^3 - 2\pi M^2)$ erg/cm$^3$, $\gamma = 2\pi \times 2.8 \times 10^6$ s$^{-1} \cdot$ Oe$^{-1}$, and $K_{cm} = -5$ erg/cm$^3$.  
		}
	}
	\label{fig:Delta_Hr}
\end{figure}
}

\section*{Discussion}
{
To unambiguously distinguish the macroscopic ICME from the localized dynamics of photoactive impurity centers, the system must be analyzed in the time domain via ultra-fast subpicosecond laser excitation. The fundamental distinction between these two competing mechanisms lies in their characteristic temporal response functions. The ICME is a non-resonant, purely coherent opto-magnetic phenomenon mediated by a virtual electronic excitation wrapper. Consequently, the effective photoinduced magnetic field generated by the ICME tracks the optical pulse profile instantaneously, turning on and off within the subpicosecond duration of the laser pulse ($\tau \sim 100$ fs). This impulsive stimulus acts as a $\delta$-functional torque that initiates coherent magnetization precession (ferromagnetic resonance) without transferring thermal energy or modifying the ground-state populations. 

In sharp contrast to ICME, the mechanism involving photoactive impurity centers (such as $\text{Pb}^{3+}/\text{Pb}^{4+}$ \cite{yoshimine2014phase} or structural defect complexes) relies on real, dissipative electronic transitions. The absorption of light drives charge transfer or reorientation among these localized centers, altering the local single-ion anisotropy energy surfaces. Because this process involves real population transfers and subsequent lattice or spin-lattice relaxation, it exhibits a characteristic finite activation rise time and a long-lived metastable decay profile that typically spans from tens of picoseconds to macroscopic timescales. By deploying subpicosecond optical pulses, the instantaneous ICME torque can be temporally separated from the delayed, cumulative changes in anisotropy induced by impurity center refills, establishing a clear pathway for isolating purely opto-magnetic phenomena.
}

\section*{Conclusion}
The study demonstrated that linearly polarized CW light leads to the shift of FMR frequency by the inverse Cotton-Mouton effect. The theoretical consideration was based on the Lagrangian formalism considering magnetization dynamics in the magnetic potential taking into account a contribution from the ICME. The latter was derived from the group theory analysis. 

The FMR frequency shift is almost linear with light intensity. It depends on the polarization of light with respect to the equilibrium magnetization state. If initial magnetization is along the in-plane external magnetic field then the maximum influence of light takes place for the light polarization either parallel or perpendicular to the equilibrium magnetization. The case when initial magnetization is out-of-plane was considered in Appendix~\ref{appendix_b}.

The dependence of the resonance frequency shift on varying system parameters was further analyzed. Two equilibrium configurations were examined: one with the magnetization lying in the film plane and another with the magnetization oriented out of the plane. For both cases, the frequencies obtained from nonlinear numerical simulations were systematically compared with those predicted by the linearized model, showing quantitative agreement.

{
Crucially, the theoretical model was extended to describe traditional field-swept FMR spectroscopy by implementing a fixed-frequency constraint. By expanding the perturbed resonance condition to first order, an explicit analytical expression was derived for the optomagnetic resonance field shift. This formulation directly maps the microscopic spin-photon coupling parameter $\omega_{cm}$ onto the displacement of the external resonance field, establishing its parametric dependence on the incident light polarization angles.
}

Finally, the theoretical predictions were confronted with available experimental results. The comparison confirmed the applicability of the developed theory and enabled the determination of the ICME constants for the studied sample.

\section*{ACKNOWLEDGMENTS}
This work was financially supported by Russian Science Foundation (Project No. 23-62-10024) in part related to electromagnetic modeling of the spin dynamics. N.I.G. and V.I.B. also acknowledge support from the Foundation for the Advancement of Theoretical Physics and Mathematics “BASIS” (Project No. 25-1-1-49-4) for analytical theoretical studies.
\begin{widetext}

\appendix

\section{Kittel formula for FMR}\label{appendix_fmr}

Kittel formula for the ferromagnetic resonance frequency 
\begin{equation}
	\omega^2_{\mathrm{fmr}} = \frac{\gamma^2}{M^2 \sin^2\theta_0} 
	\left( 
	\frac{\partial^2 U}{\partial \theta^2}\frac{\partial^2 U}{\partial \varphi^2} 
	- \left(\frac{\partial^2 U}{\partial \theta \partial \varphi}\right)^2 
	\right)_{\theta=\theta_0,\varphi=\varphi_0}
	\label{fmr_littel}
\end{equation}
is obtained from the second derivatives of the total energy $U$ \cite{suhl1955ferromagnetic,kittel1948theory, kittel1951ferromagnetic}. 
The ferromagnetic resonance frequency calculated $\omega_{\mathrm{fmr}}$ using the Kittel formula \ref{fmr_littel} coincides with the previously obtained value $\omega_r$ (Eq.~\ref{wr}). This confirms the validity of the derivation and the consistency of theoretical description of the system’s dynamics.

\section{Derivation of Cotton-Mouton energy}\label{appendix_a}

A system of iron garnet thin film is considered, where a Cartesian system is introduced: the film lies in the \(x,y\) plane and its normal coincides with the \(z\)-axis. The sample is illuminated by linearly polarized light characterized by the electric field \(\mathbf{E} = E_0 (\sin\beta \cos\alpha, \sin\beta \sin\alpha, \cos\beta)\), which is parameterized by two angles \(\alpha\) and \(\beta\). Here, \(\beta\) denotes the angle of deviation of the incident light from the normal to the film, while \(\alpha\) represents the rotation of the polarization plane relative to the \(x\)-axis.
The spherical coordinates are chosen as the polar and azimuthal angles $(M,\theta,\varphi)$ of the magnetization vector $\mathbf{M}$ (in Cartesian coordinate system $\mathbf{M}=M(\sin\theta\cos\varphi, \sin\theta\sin\varphi, \cos\theta)$). 

When analyzing a physical system, the first step is to identify the symmetry group of its Hamiltonian, that is, the set of transformations that leave the Hamiltonian invariant \cite{hamermesh2012group}. In our work, we consider a system with uniaxial anisotropy, meaning that there exists a single distinguished direction along the -axis.

The crystallographic class of type $\infty\infty$ corresponds to the symmetry class of infinitely extended cylindrical objects \cite{sirotinfundamentals}. The first $\infty$ denotes an infinite rotational axis along a certain direction, while the second $\infty$ indicates the existence of an infinite set of equivalent directions perpendicular to this axis. In other words, the object possesses cylindrical symmetry. From the tensor of the Cotton–Mouton coefficients \cite{sirotinfundamentals}, one can derive the expression \cite{hamermesh2012group} for the Cotton–Mouton energy as 
\begin{align}
    U_{CM} = a_1 (M_x^2E_x^2 + M_y^2E_y^2 +M_z^2E_z^2) +a_2\Bigl(M_x^2 & (E_z^2+E_y^2)+M_y^2 (E_z^2+E_x^2)+M_z^2 (E_x^2+E_y^2)\Bigr)+\\
    &+\frac{a_1-a_2}{2}\Bigl(M_x M_y E_x E_y + M_x M_z E_x E_z +M_z M_y E_z E_y \Bigr),\nonumber
\end{align}
where $z$ axis is directed along the easy axis of the uniaxial anisotropy. Writing equations in spherical coordinates with $\mathbf{M}=M(\sin\theta\cos\varphi, \sin\theta\sin\varphi, \cos\theta)$ and $\mathbf{E}=E_0 (\sin\beta\cos\alpha, \sin\beta\sin\alpha, \cos\beta)$, omitting constants we get 
\begin{align}
    U_{CM}=\frac{a_1-a_2}{8} M^2 E_0^2\Bigl[ \cos2\beta +\cos2\theta + 3\cos2\beta\cos2\theta + \sin^2\!\beta \sin^2\!\theta\bigr(4\cos2\alpha\cos2\varphi & + \sin2\alpha\sin2\varphi\bigr) \\
    &+\cos(\alpha-\varphi)\sin2\beta\sin2\theta\Bigr].\nonumber
\end{align}
In the case $\beta=\pi/2$, which is equivalent to the normal incidence of light (along the $z$ axis), we have 
\begin{align}
	 U_{CM}= K_{cm} \Bigl[ -2\cos2\theta+\sin^2\!\theta \bigl(4\cos2\alpha\cos2\varphi  + \sin2\alpha\sin2\varphi\bigr) \Bigr],
\end{align}
where $K_{cm} = \frac{a_1-a_2}{8}M^2 E_0^2$

\section{The case of out-of-plane equilibrium magnetization state}\label{appendix_b}
In this section, we consider the case when the equilibrium position with $\omega_{cm}=0$ of magnetization lies out of the (xy) plane. This situation is realized in the case of an easy axis, when $K_u > 2\pi M^2$, and the external magnetic field $|\mathbf{H}|<\tfrac{2K_u}{M} - 4\pi M$. The equilibrium position of magnetization is determined by $\theta_0 = \arcsin\bigl(|\mathbf{H}|/(\tfrac{2K_u}{M} - 4\pi M)\bigr)$ and $\varphi_0=0$, since we consider without limitation generality $\mathbf{H}$ along the $x$ axis.

The same linearization procedure around the equilibrium position has been performed for the case $\theta_0\not=\pi/2$. 
where $\kappa_1$, $\kappa_2$, $\kappa_3$ and $\kappa_4$ are determined under the assumption $4\kappa_1\kappa_2>(\kappa_3-\kappa_4)^2$. This result coincides with the formula \ref{wr} in the case of $\theta_0 = \pi/2$.

In this case, the general form of the equations remains the same, but the coefficients have a more  form. The angles that determine the orientation of the magnetization can be represented as $\phi = \phi_0+\phi_1$ and $\theta = \theta_0+\theta_1$ with small deviations from the ground state $\theta_1, \phi_1 \ll 1$. Taking into account terms up to the first order in $\theta_1, \phi_1$, as numerical modeling indicates that this is sufficient to accurately describe the system, we derive the Euler-Lagrange linearized equations from equations \ref{theta} and \ref{phi}
\begin{align}
	&\dot{\theta}_1\!(t) = -\kappa_1\varphi_1 + \kappa_3 \theta_1 + \kappa_5,\\
	&\dot{\varphi}_1\!(t) = \kappa_2\theta_1 + \kappa_4\varphi_1 + \kappa_6,
\end{align}
where $\kappa_1 = \gamma H_x - \omega_{cm}(\cos\theta_0\cos\alpha\sin2\beta+8\sin\theta_0\cos2\alpha\sin^2\!\beta)$, $\kappa_3 = \omega_{cm}(\sin\theta_0\sin\alpha\sin2\beta-\cos\theta_0\sin2\alpha\sin^2\!\beta)$, \\
$\kappa_5 = -\omega_{cm}(\cos\theta_0\sin\alpha\sin2\beta+\sin\theta_0\sin2\alpha\sin^2\!\beta)$, $\kappa_2 = \omega_{cm} \bigl[(-2+\cos2\theta_0)\frac{\cos\theta_0}{\sin^2\!\theta_0}\cos\alpha\sin2\beta + 2\sin\theta_0(1+3\cos2\beta-2\cos2\alpha\sin^2\!\beta)\bigr] -\omega_u\sin\theta_0+\frac{\gamma H_x}{\sin^2\!\theta}$, $\kappa_4 = \omega_{cm} (2\cos\theta_0\sin2\alpha\sin^2\!\beta + \frac{\cos2\theta_0}{\sin\theta_0}\sin\alpha\sin2\beta)$ and $\kappa_6 = \omega_{cm}\bigl[\frac{\cos2\theta_0}{\sin\theta_0}\cos\alpha\sin2\beta-2\cos\theta_0(1+3\cos2\beta-2\cos2\alpha\sin^2\!\beta)\bigr] + \omega_u \cos\theta_0-\frac{\cos\theta_0}{\sin\theta_0} \gamma H_x$. These linearized equations can be solved under the initial conditions $\theta_1(0)$ and $\varphi_1(0)$, that have the form
\begin{align}
	\varphi_1(t) \xi_1^2 = \xi_2^2 + e^{\frac{\omega_3+\omega_4}{2}t}\Bigl[\bigl(\varphi_1(0)\xi_1^2-\xi_2^2\bigr)\cos\omega_r t +\frac{\sin\omega_r t}{2\omega_r}&\bigl(2\theta_1(0)\kappa_2\xi_1^2-\varphi_1(0)(\kappa_3-\kappa_4)\xi_1^2 +\\ &\kappa_2\kappa_5(\kappa_3+\kappa_4)+\kappa_6(2\kappa_1\kappa_2-\kappa_3^2+\kappa_3\kappa_4)\bigr)\Bigr],\nonumber\\
	\theta_1(t) \xi_1^2 = -\xi_3^2 + e^{\frac{\omega_3+\omega_4}{2}t}\Bigl[\bigl(\theta_1(0)\xi_1^2+\xi_3^2\bigr)\cos\omega_r t -\frac{\sin\omega_r t}{2\omega_r}&\bigl(2\varphi_1(0)\kappa_1\xi_1^2-\theta_1(0)(\kappa_3-\kappa_4)\xi_1^2 +\\ &+\kappa_5(-2\kappa_1\kappa_2-\kappa_3\kappa_4+\kappa_4^2) + \kappa_1\kappa_6(\kappa_3+\kappa_4)\bigr)\Bigr]\nonumber\\
	&\omega_r^2 = \kappa_1\kappa_2-\frac{(\kappa_3-\kappa_4)^2}{4},
\end{align}
where $\xi_1^2 = \kappa_1\kappa_2+\kappa_3\kappa_4$, $\xi_2^2 = \kappa_2\kappa_5-\kappa_3\kappa_6$ and $\xi_3^2 = \kappa_4\kappa_5+\kappa_1\kappa_6$.

The resonant frequency for this case can be represented as an analytical function:
\begin{equation}
	\omega_r^2 = \kappa_1\kappa_2-\frac{(\kappa_3-\kappa_4)^2}{4}, 
\end{equation}

\end{widetext}
\putbib[main.bib]
\end{bibunit}




\end{document}